\documentclass{iucrjournals}

\usepackage{siunitx}
\usepackage{multirow}  

\usepackage{listings}  
\usepackage[dvipsnames]{xcolor}
\usepackage{todonotes}

\usepackage{amsmath, amssymb, amsfonts}  

\lstdefinestyle{mystyle}{
    language=Python,
    breakatwhitespace=false,         
    breaklines=true,                 
    captionpos=b,                    
    keepspaces=true,                 
    numbers=left,                    
    numbersep=5pt,                  
    showspaces=false,                
    showstringspaces=false,
    showtabs=false,                  
    tabsize=2,
    basicstyle=\ttfamily\footnotesize,
    columns=fullflexible,
    xleftmargin=2.5ex,
    backgroundcolor=\color[HTML]{fafafa},
    commentstyle=\color[HTML]{666666},
    keywordstyle=\color[HTML]{a626a4},
    numberstyle=\tiny\color[HTML]{9d9d9f},
    stringstyle=\color[HTML]{50a14f},
}
\title{HoToPy: A toolbox for X-ray holo-tomography in Python}
 
\author[a]{Jens Lucht\IUCrCemaillink{jens.lucht@uni-goettingen.de}\IUCrOrcidlink{0000-0003-4447-7882}}%
\author[a]{Paul Meyer\IUCrOrcidlink{0000-0002-9193-2000}}%
\author[b,a]{Leon Merten Lohse\IUCrOrcidlink{0000-0001-5058-7215}}%
\author[a]{Tim Salditt\IUCrCemaillink{tsaldit@gwdg.de}\IUCrOrcidlink{0000-0003-4636-0813}}

\affil[a]{Georg-August-Universität Göttingen, Institut für Röntgenphysik, Friedrich-Hund-Platz 1, 37077 Göttingen, Germany}
\affil[b]{Universität Hamburg, The Hamburg Centre for Ultrafast Imaging, Luruper Chaussee 149, 22761 Hamburg, Germany}

\begin{document}
\maketitle 


\begin{abstract}
We present a Python toolbox for holographic and tomographic X-ray imaging. It comprises a collection of phase retrieval algorithms for the deeply holographic and direct contrast imaging regimes, including non-linear approaches and extended choices of regularization, constraint sets, and optimizers, all implemented with a unified and intuitive interface.
Moreover, it features auxiliary functions for (tomographic) alignment, image processing, and simulation of imaging experiments. 
The capability of the toolbox is illustrated by the example of a catalytic nanoparticle, imaged in the deeply holographic regime at the `GINIX' instrument of the P10 beamline at the PETRA III storage ring (DESY, Hamburg). 
Due to its modular design, the toolbox can be used for algorithmic development and benchmarking in a lean and flexible manner, or be interfaced and integrated in the reconstruction pipeline of other synchrotron or XFEL 
instruments for phase imaging based on propagation.
\end{abstract}


\section{Introduction}

\begin{figure}
  \includegraphics{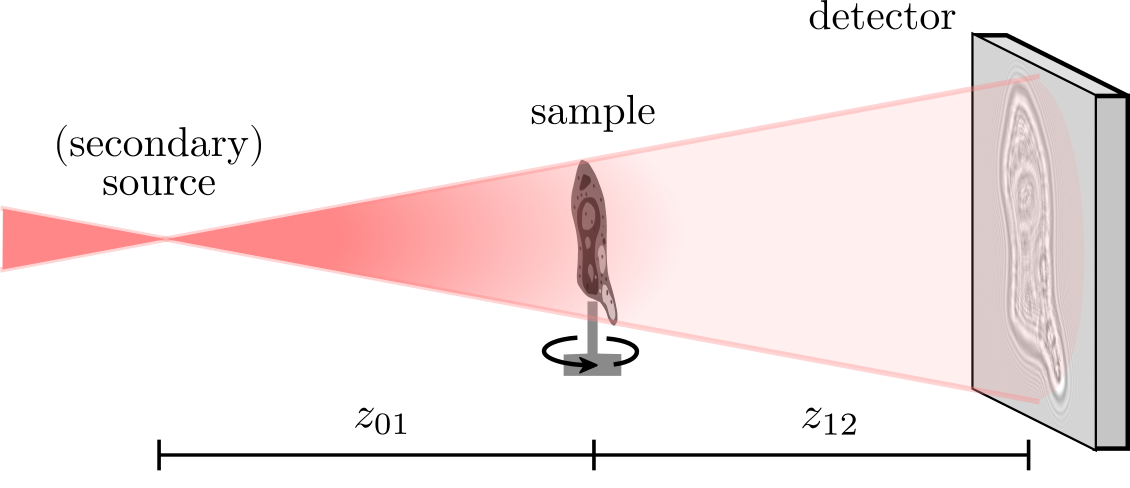}
  \centering
 \caption{Schematic for propagation-based imaging (PBI) in cone-beam geometry. A diverging X-ray beam illuminates a sample with weak absorption, which is placed at a defocus distance $z_{01}$. The sample imprints phase shifts onto the X-ray wavefront, which render into a measurable self-interference pattern on a detector, placed at distance $z_{12}$ downstream the sample, after sufficient free-space propagation. For tomography the sample is rotated and imaged at multiple angles.}
 \label{fig:pbi}
\end{figure}

The ability of X-ray radiation to penetrate matter is key to its use as a non-destructive probe for
the inner structure of objects, materials and tissues, by ways of computed tomography (CT).
Penetration relies on weak attenuation and therefore has always been a limitation as much as an
enabling property.
Contrast vanishes for soft tissues and low-$Z$ materials at
scales in the mircometer range and below, when attenuation
becomes insufficient.
The more recent exploitation of phase contrast based on (partially) coherent
beam propagation has helped to overcome these limits, offering sufficient contrast even to unstained soft
biological tissue, soft matter materials, and/or nanoscale structures in solution.
Different experimental techniques exist to transform phase shifts imparted by a sample into measurable intensity patterns. For example propagation based imaging (PBI), where intensity patterns emerge through self-interference after sufficient (optics-free) free space propagation, as illustrated in \autoref{fig:pbi}. In pytchography this is additionally combined with lateral sample shifts \cite{Rodenburg2004,Robisch2015,Pfeiffer2018pty}. Other techniques probe the phase by additional optical elements such as grating-based \cite{Momose2009,Pfeiffer2013gbi} or speckle imaging \cite{Zanette2014speckle}.
For a detailed introduction to X-ray phase contrast imaging we refer to the references \cite{Paganin2006,Salditt2020_part_holotomo,Quenot2022}.
Hence, X-ray phase contrast imaging and computed tomography
(XPCT) is an unique tool for a wide range of applications. Recent cutting edge examples are
diverse, including
nanoimaging of neuronal tissue for connectomics \cite{Livingstone_2025,Azevedo_2024},
morphological transitions of nanoparticles in solution \cite{Grote_2022,Vesely_2021}, or ultrafast
imaging of hydrodynamics at X-ray free electron lasers (XFELs) for cavitation
\cite{Hoeppe_NJoP_2024} and fusion confinement research \cite{Montgomery2023}. In all cases,
efficient and high-quality phase retrieval is a key element in phase contrast imaging, in particular
in the high resolution full-field variant of holographic tomography (holo-tomography) \cite{Cloetens_1999}.

Emerging capabilities of fourth-generation synchrotrons \cite{Chushkin2025,Li2022} and XFEL sources, of larger and faster detectors \cite{Correa2024,Donath2023eiger2,Hatsui2015}, as well as of optics and instruments enable higher spatial resolution, dose efficiency, larger fields of view, and shorter acquisition times \cite{vijayakumar2024,Spiecker_2023,Astolfo2017}. At the other end of the image chain, new paradigms in image analysis are fueled by machine learning, which requires large image libraries, for example as training data \cite{Bellens2024,Flenner2022,Hendriksen2020}. To meet both ends, reconstruction software and computing pipelines must keep up with the boost in efficiency and throughput, while at the same time achieving image quality beyond the standard linearized phase retrieval approaches.

A variety of toolboxes already exist for X-ray holography, tomography, and ptychography, reflecting the growing demand for advanced reconstruction software.
A few examples are \emph{PyPhase} \cite{Langer2021}, \emph{CIL} \cite{Joergensen2021}, \emph{TomoPy} \cite{Guersoy2014}, \emph{HoloTomoToolbox} \cite{Lohse2020}, \emph{Holotomocupy} \cite{Nikitin_2023}, \emph{TOFU/UFO} \cite{Farago_2022,Vogelgesang2016}, \emph{PyHST} \cite{Mirone2014}, \emph{PtyLab} \cite{Loetgering2023}, \emph{PyNX} \cite{FavreNicolin2020} and \emph{PtyPy} \cite{Enders2016}.
Each of these has its own strengths and capabilities.
With HoToPy, we contribute implementations of recent advances in phase retrieval such as \cite{Farago_2024,Huhn_2022}, iterative tomographic alignment techniques and image processing tools, that to our knowledge are not yet available in other frameworks. We pair this with high level implementations of smooth and non-smooth optimization.  
%
Emerging from the HoloTomoToolbox \cite{Lohse2020} for MATLAB, \emph{HoToPy} not only serves as a drop-in replacement but extends it with recent methods in a more flexible and modular framework.
By leveraging the Python ecosystem including the `PyTorch' \cite{pytorch} library for GPU acceleration and automatic differentiation, \emph{HoToPy} seamlessly integrates with other existing frameworks simplifying integration into existing analysis workflows and pipelines. 
It offers a broad spectrum of functionality for the entire range of data analysis, from image processing, phase retrieval, alignment, to tomographic reconstruction, but also the simulation of experiments. The source code is licensed under GNU General Public License and publicly available at Ref. \cite{hotopy_code}. Up-to-date installation instructions, detailed documentation and examples are provided therein.

Currently, HoToPy is rapidly evolving. In this manuscript, we showcase its
current status and demonstrate its capabilities on experimental data taken on the \emph{Göttingen
instrument for nano-imaging with X-rays} (GINIX) operated by our research group at the P10 beamline
of the PETRA III storage ring at DESY in Hamburg.

\section{HoToPy toolbox}

The HoToPy package is a Python toolbox for holographic and tomographic X-ray imaging. Its primary
use is the reconstruction of propagation based X-ray phase contrast
tomography---\emph{holo-tomography}---data recorded at synchrotron or laboratory X-ray sources,
but it can also be used for visible light or electron holographic imaging. 
HoToPy is implemented using the machine learning library `PyTorch' \cite{pytorch}, which provides strong GPU acceleration and flexible automatic differentiation.
High-performance tomographic primitives are provided through an interface to the ASTRA toolbox \cite{Palenstijn2013,Aarle2015,Aarle2016}. HoToPy can either be used as a software library in reconstruction pipelines, or it can be used interactively by the user, thanks to its sane defaults and intuitive interface. Furthermore, real experimental datasets for testing and development of novel algorithms are included.

\subsection{Numerical concepts}

HoToPy's flexibility, extensibility and high numerical performance resides on two core principles: the use of \emph{state-of-the-art numerical optimization algorithms} and \emph{automatic differentiation (AD)}, whose concepts are briefly introduced in the following.
In order to ease adoption of these concepts we provide reusable optimization algorithms through the \texttt{hotopy.optimize} submodule.

\paragraph{State-of-the-art numerical optimization} underpins HoToPy's phase retrieval algorithms and enable fast and robust reconstructions with high numerical efficiency. We provide algorithms for smooth and non-smooth optimization, including a robust \emph{proximal gradient method} (PGM) with backtracking line search and adaptive step sizes described in \cite{Goldstein_2014}, an accelerated \emph{alternating direction method of multipliers} (ADMM) \cite{Goldstein2014_admm} and \emph{fast iterative shrinkage-thresholding algorithm} (FISTA) \cite{Beck2009}.
In particular, the ADMM algorithm is used for the constrained variant of the contrast transfer function (CTF) and intensity transfer function (ICT) phase retrieval algorithms and the PGM with automatic differentiation for {Tikhonov} and {TikhonovTV}, as described in \cite{Huhn_2022}.
%

\paragraph{Automatic differentiation} (AD) is popular for training of (deep) neural networks \cite{Paszke2017,Baydin2018}, where it is used 
to dynamically compute the gradient of a scalar-valued objective or loss function. Thus, neither finite-difference approximation nor analytical derivation and explicit implementation of gradients is required. In general, this allows fast prototyping of new algorithms or adaptation of existing ones without the burden to manually derive and update explicit gradients with an insignificant overhead in performance.

In HoToPy AD is automatically used within the gradient-based optimization algorithms, e.g. PGM and FISTA, and thus in the phase retrieval algorithms based on them, such as the {Tikhonov} algorithm.
Using AD provides a high degree of flexibility. For example, this allows to extend or modify the Tikhonov algorithm with \emph{any} (smooth) regularization, such as a smoothed $L^1$-norm for a total variation (TV) regularization \cite{Hansen2021book} used in {TikhonovTV}.
Furthermore, \emph{any} modification of the forward model can easily be combined with any other constraint or regularization.

\subsection{Phase retrieval}
\label{sec:phaseretrieval}

Since the phase of an X-ray wavefront cannot be measured with current technology,
a central step in holo-tomography is the computational reconstruction of quantitative phase and attenuation images from recorded near-field diffraction patterns, or (inline) \emph{holograms}.
This so-called ``phase problem'' poses a nonlinear, ill-posed inverse problem. To solve this inverse problem, several algorithms have been developed incorporating different assumptions and priors, such as assuming short propagation distance, a single-material object \cite{Paganin_2002} or an optically weak object \cite{Cloetens_1999}. 
Generally, in PBI two regimes can be distinguished characterized by a dimension-less quantity, the \emph{Fresnel number}. It is defined by $\mathfrak{f} = \frac{\sigma^2}{z\lambda}$, with reference length scale (e.g. sample diameter, structure size or pixel size) $\sigma$, propagation distance $z$ between sample and detector and wavelength $\lambda$. One regime is the so-called \emph{direct contrast} or edge-enhancement regime, where $\mathfrak{f} \cong 1$, the other is the \emph{holographic regime}, where $\mathfrak{f} \ll 1$.
For each regime a range of different algorithms are available in HoToPy, see \autoref{tab:phase_algos}.

In HoToPy, holography related methods (propagation and phase retrieval) are provided through the \texttt{hotopy.holo} submodule. So far, it is focused on propagation-based phase contrast
in the direct-contrast as well as holographic regime. All
algorithms for the holographic regime support imposing priors as object constraints, e.g.
pixel-wise non-positivity of the phase or finite supports. An overview of the
available phase retrieval algorithms at the time of writing this article is given in
\autoref{tab:phase_algos}, while an updated list can be found in the online documentation. All are implemented with GPU computation support. Moreover, they allow for astigmatism, so that the
effective propagation distance or equivalently the Fresnel number $\mathfrak{f}$ can be different in
the two directions orthogonal to the optical axis.
This can for example easily be caused by anisotropic magnifications in horizontal and vertical direction, see for example the case
of Bragg magnifiers \cite{Spiecker_2023}. 
For a detailed documentation of the algorithms we refer to the online documentation, example notebooks and their respective literature.
In addition, a number of methods for preprocessing are provided. These include the automated removal of faulty pixels as well as principal-component-analysis-based \cite{Nieuwenhove2015} and curvature-based methods for empty-beam division.

\begin{table}[htb]
    \centering
    \begin{tabular}{lll}
        \toprule
         Name & Class & Reference \\ \midrule
         \multicolumn{3}{l}{\emph{Holographic regime}} \\ 
         \cmidrule{1-1}
         Contrast transfer function (CTF) & \texttt{CTF} & \protect\cite{Cloetens_1999} \\
         \quad --- constrained CTF & \texttt{CTF} & \protect\cite{Huhn_2022} \\
         Intensity transfer function (ICT) & \texttt{ICT}  & \protect\cite{Farago_2024} \\
         nonlinear Tikhonov & \texttt{Tikhonov} & \protect\cite{Huhn_2022} \\
         \quad --- (smoothed) TV regularization & \texttt{TikhonovTV} & (Lucht \textit{et al., unpublished}) \\
         Alternating Projections (AP) & \texttt{AP} & \protect\cite{Hagemann_2018} \\ 
\addlinespace
         \multicolumn{3}{l}{\emph{Direct contrast regime}} \\ 
         \cmidrule{1-1}
         Paganin & \texttt{Paganin} & \protect\cite{Paganin_2002} \\
         generalized Paganin & \texttt{GeneralizedPaganin} & \protect\cite{Paganin_2020} \\
         Bronnikov-aided correction & \texttt{BronnikovAidedCorrection} & \protect\cite{DeWitte_2009} \\
         modified Bronnikov & \texttt{ModifiedBronnikov} &  \protect\cite{Groso_2006} \\
         \bottomrule
    \end{tabular}
    \caption{Overview of available phase retrieval methods for the holographic as well as the direct contrast regime.}
    \label{tab:phase_algos}
\end{table}

\subsection{Computed tomography}

The tomographic methods are organized in the \texttt{hotopy.tomo} submodule. Through interfaces to the ASTRA toolbox \cite{Palenstijn2013,Aarle2015,Aarle2016}, efficient (multi-)GPU reconstruction and projection algorithms in two and three dimensions are provided, including e.g. filtered back-projection (FBP), the cone-beam algorithm of Feldkamp, Davis, and Kress (FDK) and the Simultaneous Iterative Reconstruction Technique (SIRT).
For both parallel beam and cone beam geometric models, the source, sample, and detector can be positioned freely for each projection image. Thus, any inexactness of the tomographic trajectory can be incorporated directly into the geometric model. This is computationally more efficient than aligning and interpolating the projection images which also degrades image quality. The toolbox contains methods based on image registration \cite{Guizar_2008} for determining deviations from an assumed tomographic trajectory.

The center of rotation (CoR) can either be found by registering the shift between two opposing projection images, or, for scans with an angular range larger than \SIrange{0}{180}{\degree}, two opposite segments of a sinogram---ideally two half rotations---can instead be registered. In the latter, the CoR estimate becomes an average over all acquisition angles, making it more robust for acquisitions with little position precision. The registration can be repeated for sinograms from different detector rows to further increase robustness and also determine small tilts of the rotation axis.

The iterative \emph{reprojection alignment} algorithm \cite{Leeuwen_2018} in the toolbox enables reconstructing and correcting rigid sample movement between individual projections during a tomographic scan. 
In each iteration, the volume is reconstructed based on the current geometry estimate. (Re-)Projection images are generated from the reconstructed volume, registered against the acquired images and the geometric model for the respective projection updated according to the detected shift. In practice it is often advisable to apply pixel binning and a bandpass filter to the projection images prior to the alignment routine to accelerate the computation and improve the registration.
Nonlinearities in the detector response, but also illumination can cause stripes in the
sinograms which lead to ring artifacts in the reconstructed volumes. HoToPy contains implementations
of additive \cite{Ketcham_2006} and wavelet based \cite{Munch_2009} ring-removal algorithms to
mitigate these artifacts.

\section{Reconstruction example: Catalytic particle}

We demonstrate the HoToPy toolbox with the example of an X-ray holo-tomography dataset of an isolated
 catalytic particle with a diameter of approximately \qty{33}{\um} used for olefin
polymerization. For an in-depth description of the sample, and how the inner structure matters for
the cataytic function in the application context, we refer to
\cite{Werny_2022,Werny_2022_jacs,Vesely_2021}.
The particle is attached to the interior wall of a Kapton tube. The particle morphology and fragmentation (`cracks') is of particular interest.
The data was recorded at the P10 beamline of the PETRA III storage ring at DESY in Hamburg with the 
\emph{Göttingen Instrument for Nano-Imaging with X-Rays} (GINIX). The GINIX was used in its cone-beam configuration to achieve geometrical magnification and effective pixel sizes down to the nanometer range. To this end, the incident beam is focused with a Kirkpatrick–Baez (KB) mirror system onto an X-ray waveguide acting as quasi-point source for holographic illumination.
The waveguide (ID 4743) is a combination of two orthogonally crossed thin-film waveguides with a diameter of $\qty{58}{\nm}$ and depth of $\qty{600}{\um}$ each, functioning together as a two dimensional waveguide \cite{Krueger_OE_2010}.
Two tomograms at two defocus distances were acquired with 1501 projections each, covering an angular range of $\ang{180}$. The source-to-sample distance $z_{01}$ was adjusted to $\qtylist{13.53; 16.73}{\mm}$, at  constant source-to-detector distance $z_{02} = \qty{5110}{\mm}$. 
The photon energy $E_\mathrm{ph}=\qty{13.8}{\keV}$ was selected by a Si(111)-monochromator.
Images were recorded using a Gadox scintillator of \qty{15}{\um} thickness fibre-coupled to an Andor
Zyla sCMOS sensor with a pixel size $\Delta_x$ of \qty{6.5}{\um} and $2160\times2560$ pixels.
The exposure time per acquisition was \qty{1}{\s.} The geometry corresponds to a geometric
magnification of $M = z_{02}/z_{01} = \num{378}$, respectively $\num{305}$, resulting in an
effective pixel size of $\qtylist{17.2; 21.3}{\nm}$. The Fresnel numbers (with respect to the effective pixel
size), $\mathfrak{f} = \Delta_x^2/[M \lambda (z_{02} - z_{01})]$, evaluate to $\numlist{2.44e-4;
3.02e-4}$.

\paragraph{Phase retrieval}

First, the phase of the catalytic particle has to be reconstructed by phase retrieval. To this end,
the recorded raw intensity patterns are preprocessed by dark current subtraction and divided by
interpolated empty beam images, i.e. images taken without a sample. The empty image $\imath$ used for
normalization at the tomographic angle $\theta$ with range $[0,
\theta_\mathrm{max}]$ is interpolated by a linear combination of the (average) empty images
before $\imath_\mathrm{pre}$ and after $\imath_\mathrm{post}$ the tomographic scan,
$
\imath(\theta) = \imath_\mathrm{pre}\left(1 - \theta / \theta_\mathrm{max}\right) +
\imath_\mathrm{post}\theta / \theta_\mathrm{max}.
$
The resulting normalized holographic diffraction patterns are also called (inline)
\emph{holograms}. Afterwards, residual low frequency background variations are suppressed by a least
curvature inpainting of the background within a compact support of the particles. Prior to phase
retrieval the holograms of the second distance are magnified to a effective parallel beam geometry
with effective pixel size $\qty{17.2}{\nm}$ and respective Fresnel numbers $\numlist{2.44e-4;
1.98e-4}$.

\begin{figure}[ht] %
\centering
    \includegraphics{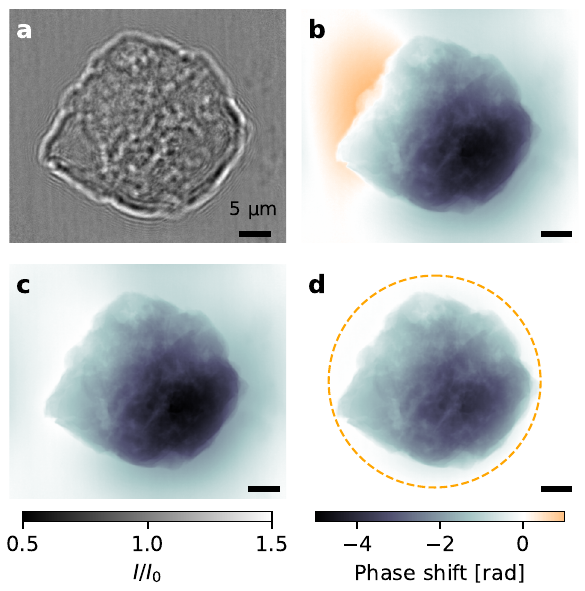} 
    \caption{Hologram and phase reconstructions of a catalytic particle. \textbf{a} Exemplary shows one of the two normalized holographic intensity interference patterns $I/I_0$ (hologram) of a catalytic particle at one tomographic angle.
    \textbf{b}-\textbf{d} Comparison of different phase retrieval methods and constraints. The reconstruction in \textbf{b} uses an unconstrained linear contrast transfer function (CTF). The reconstructions \textbf{c} and \textbf{d} are obtained using the HoToPy--Tikhonov algorithm. For these, a pixel-wise non-positivity constraint is used and for \textbf{d} additionally a finite disk-shaped support, indicated by the dashed circle. Scale bars: $\qty{5}{\um}$. Effective pixel size: $\qty{17.2}{\nm}$. Images have $2160\times2560$ pixels.
    }
    \label{fig:cat_holo}
\end{figure}

An exemplary empty-beam divided hologram is shown in \autoref{fig:cat_holo}a together with different phase reconstructions in \autoref{fig:cat_holo}b-c. \autoref{fig:cat_holo}b shows a single step phase reconstruction using the contrast transfer function (CTF) method \cite{Cloetens_1999} without the use of any constraints. \autoref{fig:cat_holo}c-d show a non-linear and constraint-based phase reconstruction using the \texttt{Tikhonov} \cite{Huhn_2022} algorithm with pixel-wise non-positivity, and non-positivity combined with a compact disk-shaped support, respectively. All reconstructions were computed with the input of the recordings of the two distances, assuming a homogeneous object with beta-delta-ratio $\beta/\delta = 0.035$
and applying a two-level frequency regularization using the weights $\alpha_\mathrm{low} = \num{2e-5}, \alpha_\mathrm{high} = \num{3e-5}$ \cite{Huhn_2022}.

By comparing the reconstructions, we can directly observe strong background variations in \autoref{fig:cat_holo}b and c. In the lower right corner of \autoref{fig:cat_holo}c we can observe that this also affects the phase reconstruction within the particle. Furthermore, the linear and unconstrained reconstruction in \autoref{fig:cat_holo}b incorrectly contains positive values, due to the convention used here a higher density sample relative to the empty beam should solely have non-positive values (negative phase shift or retarded waves). Thus, using this as prior knowledge for the reconstructions yields more faithful reconstructions (\autoref{fig:cat_holo}c). Finally, if additionally combined with a disk-shaped support, low frequency background variations are effectively suppressed (\autoref{fig:cat_holo}d).

The corresponding code snippets used for the phase reconstructions are given in
\autoref{code:catpart_phase}. A comparison of computation times of two phase retrieval
algorithms with different sets of constraints is given in table \autoref{tab:phase_bench}.

\paragraph{Tomography}

\begin{figure}[ht]
    \centering
    \includegraphics{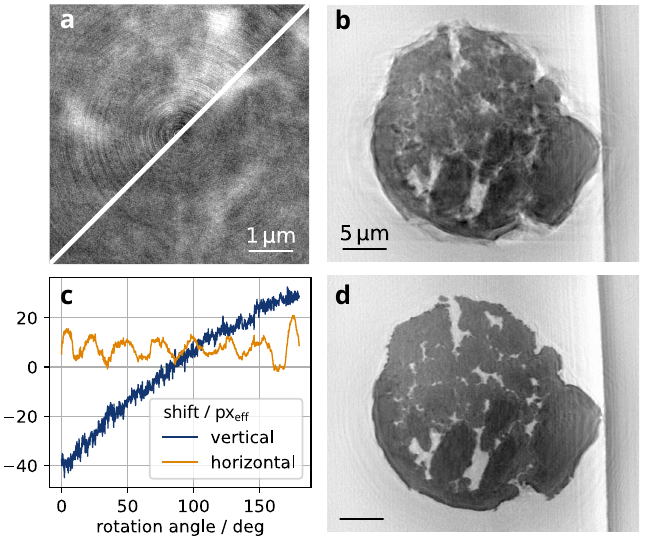}
    \caption{Tomographic reconstruction by the FDK algorithm. \textbf{a} shows a zoom into the center of a reconstructed horizontal slice. Concentric ring artifacts (top left) are mitigated (bottom right) after applying wavelet-based ring removal ($l=4$, $\sigma=1$) to the sinogram. \textbf{b} displays a vertical slice through a volume reconstruction assuming  an idealized acquisition trajectory. \textbf{c} shows the shifts of the projection images estimated by registration of opposite projections and reprojection registration ($8\times8$ pixel binning, high-pass filter with $\sigma=5$, 100 iterations). \textbf{d} shows a virtual slice through the volume reconstruction after applying the shift correction. The effective voxel size is $\qty{17.2}{nm}$.}
    \label{fig:cat_tomo}
\end{figure}

A three dimensional model of the sample is created from the phase projections by means of
tomographic reconstruction. A wavelet-based filter \cite{Munch_2009} is applied to reduce stripes in the sinograms.
The corresponding reduction of ring artifacts in the reconstructed slice can be seen in
\autoref{fig:cat_tomo}a. The direct tomographic reconstruction of the sinograms
(\autoref{fig:cat_tomo}b) suffers from artifacts caused by inexactness of the acquisition
trajectory, mainly due to sample movement. With a geometric model which includes the sample movement
(\autoref{fig:cat_tomo}c), the quality of the reconstruction improves drastically and fragmentation
cracks can be traced (\autoref{fig:cat_tomo}d). To extract the sample movement from the acquired
projection images, the CoR and drift were first estimated by registration of opposite projections.
The reprojection alignment algorithm was then applied to a reduced dataset which was obtained by
$8\times8$ pixel binning and high-pass filtering with a Gaussian filter kernel ($\sigma = 40/8$).
The influence of the Kapton tube's strong edge was reduced by a directional Fourier filter applied
to images with observation angles close to the edge surface. The displayed shifts were obtained
after 100 iterations, taking about \qty{9.3}{\min} on a machine with 24 CPU cores (Intel(R) Xeon(R)
w7-3455) and an NVIDIA RTX 6000 Ada Generation GPU. The FDK
reconstruction of the full volume takes about \qty{2}{min}. The code snippets for the tomography
reconstruction are given in \autoref{code:cat_tomo}.

\section{Conclusion}

We have presented \emph{HoToPy}, an open source toolbox for holographic and tomographic X-ray imaging in Python. It provides a collection of phase retrieval algorithms for the direct contrast and in-line holographic regime, suitable for propagation-based phase imaging at synchrotron and laboratory $\mathrm{\mu}$CT instruments alike. Tomographic reconstruction and alignment methods are also included. It is easily extensible without sacrificing speed by leveraging automatic differentiation and GPU computation, facilitating rapid testing of novel algorithms.
Compared to our current `workhorse' the \emph{HoloTomoToolbox} for MATLAB, \emph{HoToPy} not only integrates into the extensive ecosystem of scientific computing in Python but also features iterative tomographic reprojection alignment and new phase retrieval algorithms such as ICT \cite{Farago_2024} and TikhonovTV, extending the approach of \cite{Huhn_2022} to TV regularization.
We have showcased the toolbox by reconstructing the volume from a deeply holographic and severely misaligned dataset of a catalytic particle.

The toolbox can serve at least four different purposes: First, it can be used for testing and development of algorithms, operated stand-alone on pre-recorded or simulated data, for example using the supplied phantoms. Second, in the same manner it can serve class-room teaching and
visualization. Third, it can be operated with a script-based analysis pipeline for a specific instrument such as the `{GINIX}' instrument.
In this case, wrapper scripts handling the instrument specific metadata and data formats, or default reconstruction parameters can easily be written and supplied to the user.
Fourth, algorithmic implementations and functions of interest can be integrated into entirely different reconstruction pipelines tailored to specific instruments and requirements.  
For example, integration and interoperability with other phase and tomographic reconstruction platforms such as  CIL \cite{Joergensen2021} should be possible with little effort.  Finally, beyond its primary use as a tool for X-ray phase imaging, we envision it to be useful also for inline optical or electron holography. To this end, the generic formulation in units of pixel size, Fresnel number or real and imaginary part of the refractive index is particularly helpful.  

\appendix 
\section{Code examples from HoToPy}

For the complete reconstruction script, full documentation and further examples, please refer to Ref. \cite{hotopy_code}. The given code examples in this manuscript are based on HoToPy version v0.21.

\subsection{Reconstruction example: catalytic particle}
\begin{lstlisting}[language=Python, caption=Source code snippet for phase reconstruction of a catalytic particle with HoToPy shown in \autoref{fig:cat_holo}., label=code:catpart_phase]
from hotopy.holo import CTF, ICT, Tikhonov, Constraints

# [...] data loading and preprocessing

# phase retrieval parameters
shape = holograms.shape[-2:]
betadelta = 0.0035  # single material object
alpha = (2e-5, 3e-5)  # frequency regularization parameters
device = "cuda:0"  # torch.device string

# setup of phase retrieval algorithms
ctf = CTF(shape, fresnel_numbers, betadelta=betadelta, alpha=alpha, device=device)
tikhonov = Tikhonov(shape, fresnel_numbers, betadelta=betadelta, alpha=alpha, device=device)
ict = ICT(shape, fresnel_numbers, betadelta=betadelta, alpha=alpha, device=device)

# pixel-wise non-positivity constraint
c_nonpos = Constraints(phase_max=0.0)

# Figure b: linear CTF without constraints
rec_ctf = ctf(holograms).cpu()

# not shown: alternatively ICT without constraints
rec_ict = ict(holograms).cpu()

# not shown: CTF with non-positivity constraint
rec_ctf_nonpos = ctf(holograms, constraints=c_nonpos).cpu()

# Figure c: nonlinear reconstruction with non-positivity constraint
rec_tikhonv = tikhonov(holograms, constraints=c_nonpos).cpu()

# Figure d: same as above but additionally with support mask
# generate mask, where ball is non-zero
from hotopy.image import ball
radius = 980  # pixel
center = (shape[0]/2-0.5, shape[1]/2-50)
mask = ball(shape, radius, center=center) > 0  # binary mask

c_nonpos_support = c_nonpos * Constraints(support=mask)  # concatenation is supported
rec_tikhonov_support = tikhonov(holograms, constraints=c_nonpos_support).cpu()
\end{lstlisting}

\begin{lstlisting}[language=Python, caption=Source code snippet for the tomographic reconstruction and reprojection alignment to correct the acqusition trajectory shown in \autoref{fig:cat_tomo}., label=code:cat_tomo]
from hotopy import tomo, image
import numpy as np

# [...] load `projections` and define directional filter `filter_vertical_edge`
#       which reduces artifacts caused by the strong kapton edge

# ringremoval
projections = tomo.ringremove(projections, type="wavelet", level=4, sigma=1)

# pixel binning
binning = 8
proj_small = image.AveragePool2d(binning)(projections)

# image filter
proj_small = filter_vertical_edge(proj_small)
sigma_low, sigma_high = 0, 40 / binning
proj_small = image.GaussianBandpass.apply(proj_small, sigma_low, sigma_high, pad="auto", ndim=2)
proj_small = proj_small[:,20:-20, 20:-20]  # crop borders

# CoR/drift estimation and reprojection alignment
shift, _ = image.register_images(proj_small[0], np.fliplr(proj_small[-1]).copy())
initial_shifts = np.c_[shift[1] / 2 * np.ones(len(angles)),             # horizontal
                       shift[0] / 2 * np.linspace(1, -1, len(angles))]  # vertical
z01, z02, px = 13.53e-3, 5110e-3, 6.5e-6  # in m
t = tomo.setup(proj_small.shape[-2:], angles, cone=(z01, z02, px * binning), sino_pad="corner")
rep_align = tomo.ReprojectionAlignment(t, proj_small)
rep_shifts = rep_align(initial_shifts, tol=0.01, max_iter=100, upsample=100) * binning

# reconstructions
t = tomo.setup(projections.shape[-2:], angles, cone=(z01, z02, px), sino_pad="corner")
direct_reconstruction = t.reconstruct(projections)
t.apply_shift(rep_shifts)
improved_reconstruction = t.reconstruct()
\end{lstlisting}

\section{Computations times for phase retrieval}
\label{sec:benchmarks}
We performed a comparison of the computation times for the phase retrieval listed in
\autoref{tab:phase_bench}.
Here we compared the CTF and HoToPy--{Tikhonov} algorithms with and without enforcing pixel-wise non-positive phase. Additionally, for the {Tikhonov} also a disk-shaped object support was tested. The single-angle reconstructions were performed on the dual distance holograms shown in \autoref{fig:cat_holo} ($2160\times2560$ pixels), which corresponds to a single angle of the tomogram. All computations were done with single precision floats. We used default settings in HoToPy with homogeneous object assumption with a beta-delta-ratio $\beta/\delta = 0.0035$ and two-level frequency regularization $\alpha_\mathrm{low} = \num{2e-5},
\alpha_\mathrm{high} = \num{3e-5}$. The iterations were stopped with the default stopping conditions. For the single tomographic angle reconstruction the iterations
needed to met these condition are given. The phase reconstruction for the full dataset were performed sequentially for each tomographic angle on the GPU.

Clearly, the linear CTF algorithm without constraints is the fasted in the comparison, since it is a single-step algorithm without iterations. When constraints are set, like the pixel-wise phase non-positivity used here, ADMM-iterations are applied which takes approximately one order of magnitude longer than the single-step method. The HoToPy--Tikhonov algorithm takes again one order of magnitude longer to converge without constraints. If pixel-wise phase non-positivity is added, convergence is reached in a comparable time. This time is approximately doubled if phase non-positivity is combined with a disk-shaped support.

This benchmark gives an impression of the performance HoToPy's phase retrieval using a real dataset on standard hardware. When considering these results, one should keep in mind, that the performance also strongly depends on the dataset, parameters and constraints.
\begin{table}[htbp]
    \centering
    \begin{tabular}{lll}
        \toprule
         Algorithm & RTX 6000 Ada & H100 \\ \midrule
         \multicolumn{3}{l}{\emph{Single tomographic angle, two distances}} \\ 
         \cmidrule{1-1}
         CTF, no constraints (\emph{single-step method}) & \qty{4.1}{ms} & \qty{3.9}{ms} \\
         CTF, phase $\phi \leq 0$ (\emph{24 iterations}) & \qty{38.5}{ms} & \qty{30.2}{ms} \\
         \texttt{Tikhonov}, no constraints (\emph{25 iterations}) & \qty{245.8}{ms} & \qty{217.9}{ms} \\
         \texttt{Tikhonov}, phase $\phi \leq 0$ (\emph{25 iterations}) & \qty{284.5}{ms} & \qty{233.8}{ms} \\
         \texttt{Tikhonov}, phase $\phi \leq 0$ \& disk-shaped support (\emph{30 iterations}) & \qty{439.5}{ms} & \qty{317.6}{ms} \\
    \addlinespace
         \multicolumn{3}{l}{\emph{Full dataset of catalytic particle}} \\ 
         \cmidrule{1-1}
         CTF, no constraints & \qty{9}{s} & \qty{6}{s} \\
         CTF, phase $\phi \leq 0$ & \qty{72}{s} & \qty{47}{s} \\
         \texttt{Tikhonov}, no constraints & \qty{6.1}{min} & \qty{4.4}{min} \\
         \texttt{Tikhonov}, phase $\phi \leq 0$ & \qty{7.4}{min} & \qty{5.1}{min} \\
         \texttt{Tikhonov}, phase $\phi \leq 0$ \& disk-shaped support & \qty{11.8}{min} & \qty{7.3}{min} \\
    \end{tabular}
    \caption{Compute times of two phase retrieval algorithms implemented in HoToPy with
different sets of constraints. Computations were performed on a NVIDIA RTX 6000 Ada respectively NVIDIA H100 GPU with the same dataset shown in \autoref{fig:cat_holo} respectively the full catalytic particle dataset.}
\label{tab:phase_bench}
\end{table}

\begin{acknowledgements}
The authors would like to thank Simon Huhn for inspiring discussions and his help analyzing the dataset,
as well as Jan Goeman and Markus Osterhoff for their kind support of our in-house computing infrastructure.
Moreover, we thank Maximilian J. Werny, Florian Meirer and Bert Weckhuysen for providing the catalytic particle.
We acknowledge Deutsches Elektronen-Synchrotron (DESY) (Hamburg, Germany), a member of the Helmholtz Association HGF, for the provision of experimental facilities. Parts of this research were carried out at PETRA III and we would like to thank Michael Sprung and Fabian Westermeier for assistance in using the P10 beamline. Beamtime was allocated for proposal II-20211052.
This research was supported in part through the Maxwell computational resources operated at DESY.
The authors are members of the Max Planck School of Photonics.
\end{acknowledgements}

\begin{funding}
We acknowledge partial funding by Max Planck School of Photonics as well as Deutsche
Forschungsgemeinschaft (DFG) (432680300 SFB 1456), and the German Minstry of Research 
and Technology for grant Holo-Tomogaphy (05K22MG1) and RECOX (05K25MG1) within the ErUM-Pro funding line.
\end{funding}

\ConflictsOfInterest{The authors declare no conflicts of interest.
}

\DataAvailability{
The source code is openly available under the GPLv3 license at Ref. \cite{hotopy_code} and the data used for the examples can be obtained at Ref. \cite{cat_dataset}.
}

\bibliography{refs} 

\end{document}